\documentstyle[12pt,epsf]{article}

\newcommand{\rf}[1]{(\ref{#1})}
\newcommand{\bea}{\begin{eqnarray}}
\newcommand{\eea}{\end{eqnarray}}

\newcommand{\e}{\mbox{e}}
\renewcommand{\d}{\mbox{d}}
\newcommand{\g}{\gamma}

\renewcommand{\l}{\lambda}

\renewcommand{\b}{\beta}
\renewcommand{\a}{\alpha}
\newcommand{\n}{\nu}  
\newcommand{\m}{\mu}

\newcommand{\del}{\delta}

\newcommand{\k}{\kappa}

\newcommand{\oh}{\frac{1}{2}}

\newcommand{\ra}{\right\rangle}
\newcommand{\la}{\left\langle}
\newcommand{\vev}[1]{\langle {#1} \rangle}

\newcommand{\mi}{\!-\!}
\newcommand{\equ}{\!=\!}

\def\void{}
\def\labelmark{}

\newenvironment{formula}[1]{\def\labelname{#1}
\ifx\void\labelname\def\junk{\begin{displaymath}}
\else\def\junk{\begin{equation}\label{\labelname}}\fi\junk}%
{\ifx\void\labelname\def\junk{\end{displaymath}}
\else\def\junk{\end{equation}}\fi\junk\labelmark\def\labelname{}}

{\ifx\void\labelname\def\junk{\end{array}\end{displaymath}}
\else\def\junk{\end{array}\right.\end{equation}}
\fi\junk\labelmark\def\labelname{}\def\junk{}
}

\newcommand{\beq}{\begin{formula}}
\newcommand{\eeq}{\end{formula}}
\newcommand{\beqv}{\begin{formula}{}}

\begin{document}

\hfill NBI-HE-99-51

\hfill TPJU-7/99

\hfill VERSION 48.2

\hfill 13 July 1999

\begin{center}
\vspace{24pt}
{ \Large \bf Abelian gauge fields coupled to simplicial quantum gravity}

\vspace{48pt}

{\sl J. Ambj\o rn,}$\,^{a,}$\footnote{email ambjorn@nbi.dk}
{\sl K. N. Anagnostopoulos}$\,^{a,}$\footnote{email konstant@nbi.dk.~~
Address after September 1st: Department of Physics, University of Crete,
P.O.Box 2208, 710 03 Heraklion, Crete, GREECE }
and
{\sl J. Jurkiewicz}$\,^{b,}$\footnote{email jurkiewi@thrisc.if.uj.edu.pl}

\vspace{24pt}

$^a$~The Niels Bohr Institute, \\
Blegdamsvej 17, DK-2100 Copenhagen \O , Denmark, 

\vspace{24pt}
$^b$~Institute of Physics, Jagellonian University, \\
ul. Reymonta 4, PL-30 059, Krak\'{o}w 16, Poland

\vspace{36pt}

\end{center}

\vspace{1.5cm}

\begin{center}
{\bf Abstract}
\end{center}

\vspace{12pt}
\noindent
We study the coupling of Abelian gauge theories to four-dimensional
simplicial quantum gravity. The 
gauge fields live on dual links. This is the correct formulation
if we want to compare the effect of gauge fields on geometry with  
similar effects studied so far for scalar fields. It shows that 
gauge fields couple equally  weakly to geometry as scalar fields,
and it offers 
an understanding of the relation between measure factors 
and Abelian gauge fields observed so-far.


\newpage

\section{Introduction}

A path integral formulation of quantum gravity includes in its
simplest version an integration over all four-dimensional geometries
of a given fixed topology, the weight of each geometry being the
Boltzmann weight of a suitable action. An attempt to implement such a
path integral representation is known as {\it dynamical
triangulation}. The integration over geometries is approximated by a
summation over all triangulations constructed from equilateral
simplices of side-length $a$, and the continuum limit is obtained by
letting $a \to 0$ for a suitable choice of bare coupling constants in
the action \cite{aj,am}.

This model has been analyzed in a series of papers, both with and
without coupling to scalar fields. Originally the hope was that an
observed second order phase transition could be used to define a
non-perturbative theory of quantum gravity, but a closer examination
revealed that the transition was (weakly) first order \cite{bbkp}.
Motivated by an effective theory of quantum gravity, which showed that
the gauge fields coupled very strongly to the infrared sector of
gravity \cite{amm}, the coupling between Abelian gauge fields and
simplicial quantum gravity was introduced and studied using the
non-compact version of the gauge field \cite{bbkptt}.  We shall refer
to this version of the model as AGM (Abelian gauge model).  Indeed,
for the first time a significant coupling between matter and gravity
was observed, and a new phase, named the ``crinkled'' phase was
observed. It appeared to be different from the so-called ``crumpled''
and ``elongated'' phases observed that far, and which both seemed
irrelevant for a continuum limit of quantum gravity. The crinkled
phase had a negative entropy exponent $\g$ and a fractal dimension
$d_{H} \approx 4$. However, it was later shown that most of the
properties of this new phase could be obtained by a simple change in
the measure of the path integral \cite{bbkptt1}, and since such a
change is of ultra-local nature it is unlikely to provide a faithful
representation of the ideas relating the trace anomaly to the infrared
behaviour of quantum gravity, assuming conformal invariance.

While the change in measure seems to capture most of the 
gravitational physics associated with the coupling of 
Abelian matter fields to simplicial quantum gravity,
the two theories are  of course not equivalent, as emphasized 
in \cite{bbkptt,bbkptt1}. One can explore 
the full consequences of the Abelian gauge fields coupled to 
simplicial quantum gravity by Monte Carlo simulations. 
Such studies were performed in 
\cite{bbkptt} and a few ambiguities as to the nature of 
the transition to the crinkled phase and of the crinkled 
phase itself remained. For this reason and as a matter of 
principle it is valuable to have an independent simulation 
of the system. 

In the non-compact formulation the Abelian gauge action is
Gaussian. There exist well-known methods relating low-dimensional
Gaussian theories on a direct lattice to their {\it dual} versions.
The methods developed on regular lattices work equally well on
dynamical lattices and we shall make use of the concept of duality to
relate theories defined on direct and dual lattices. This relation
will highlight why the originally chosen way of coupling gauge fields
to geometry produced results almost identical to a simple measure
term. It will show that this is the result of a somewhat unfortunate
choice of action, and that the ``real'' Abelian degrees of freedom
couple to simplicial quantum gravity more or less as ordinary massless
scalar fields, i.e.\ they do not influence the phase structure of pure
simplicial quantum gravity, at least for the number of independent
Abelian fields used so far.  From this point of view the crinkled
phase observed so far is {\it entirely} due to the insertion of an
ultra-local measure factor. Further, the way we implement the coupling
between the gauge fields and geometry allows us to introduce new
effective actions which might change the phase diagram of matter
coupled to gravity.

Another possibility, not studied in this paper would be to go to the
compact formulation of the gauge action. In this case, at least in
principle, one may expect a phase transition between the confined and
deconfined phases. Numerical results for the three-dimensional case
suggest however that such a transition disappears, at least for a
single $U(1)$ field \cite{renken}.  Whether some new physics may
emerge for more fields or for the non-Abelian gauge fields remains an
open problem.

\section{The model}\label{themodel}
Let $T$ be an abstract triangulation of a four-dimensional manifold.
Let $T^{*}$ be the simplicial complex dual to $T$ and $N_{d}$ denote
the number of $d$-dimensional sub-simplices in $T$.  $N_4$ is the
total volume of the manifold.  We shall compare two possible methods
of introducing the (non compact) Abelian gauge fields on such a
manifold.

The first possibility, which we call the $T$ implementation, would be
to put the (complex) matter fields at the vertices in $T$ and the
$U(1)$ gauge potentials $A_{IJ}=-A_{JI}$ on the (oriented) links (IJ)
of the triangulation $T$, with $I$, $J$ numbering the vertices of the
triangulation. At each vertex I we can perform the gauge
transformation $\chi_I$ and the gauge fields transforms as
\beq{ad1i**}
A_{IJ} \to A_{IJ}+\chi_I-\chi_J.
\eeq
Let us now choose a spanning tree in the complex $T$, i.e.\ a
connected graph of links with no loops. The number of links in the
spanning tree is $N_{0}-1$. Making use of the gauge transformation
\rf{ad1i**} we can perform the gauge fixing of all the gauge
potentials on the spanning tree, reducing them to zero.  The resulting
number of degrees of freedom is $N_1-N_0+1$.  In principle there is
nothing wrong with this implementation where the matter fields live on
the vertices. However, the number of the degrees of freedom both of
the matter fields and the gauge fields has no simple relation to the
volume of the manifold.  For the scalar fields this number is
proportional to the number of vertices $N_{0}$ (and not $N_{4}$),
which can grow as $N_{4}^{\a}$, where $\a \le 1$. In cases where one
has a sensible thermodynamic limit it should not matter which way the
coupling between matter and gravity is done, but since
``pathological'' triangulations can occur, and even dominate for some
choices of gravitational coupling constants, this choice of
implementation may have troubles with defining the extensive
quantities, like the average action etc.

The alternative possibility, which we call the $T^*$ implementation,
is to choose the $U(1)$ gauge potentials $A^*_{ij}$ to live on the
oriented links $l^*_{ij}$ of the dual graphs, connecting simplices
with labels $i$ and $j$.  Each dual link corresponds to an interface
between the neighbouring simplices.  As usual $A^*_{ji} = -A^*_{ij}$
and there are $N_3={5\over 2} N_4$ such fields.  This formulation is
in accordance with most simulations of matter fields coupled to
gravity. The matter fields are usually associated with the
four-simplices (are located in the ``centers'' of the four-simplices)
and their derivatives are calculated as differences formed between the
values of the fields in neighbouring four-simplices, i.e. they are
associated with the dual links.  The reason for choosing such an
implementation, apart from being convenient in numerical simulations,
is that the naive counting of field degrees of freedom will be
proportional to the four-volume $N_{4}$.  This property holds also for
the gauge degrees of freedom.  At each simplex we can perform a gauge
transformation $\chi^*_i$ and the gauge fields transform as
\beq{ad1*}
A^*_{ij} \to A^*_{ij}+\chi^*_i-\chi^*_j.
\eeq
Let us now choose a spanning tree in the complex $T^*$.  The number of
links in the spanning tree is $N_{4}-1$. One can gauge $A^*_{ij}=0$
for all links in the spanning tree. This leaves us with
$N_{3}-N_{4}+1= {3\over 2}N_4 +1$ physical degrees of freedom for the
gauge field. The nice feature is that this number depends only on the
volume $N_4$.

In both implementations the independent degrees of freedom can be
parametrized by the gauge invariant (oriented) plaquette variables.
In the $T$ implementation these are oriented triangles $t_{IJK}$ and
the plaquette variables satisfy:
\beq{plaq}
P(t_{IJK})= A_{IJ} + A_{JK} + A_{KI}.
\eeq 
The plaquettes are not independent variables. Because of the Abelian
nature of the gauge fields $P(t_{IJK})$ are unchanged under the cyclic
permutation of $I,J,K$ and change sign when the orientation of the
plaquette is inversed. For each triangulation one has to define the
positive orientation for the plaquette variables to avoid the double
counting.  By construction the flux through any closed two-dimensional
surface is zero. The number of independent such surfaces is (in case
the topology is that of the 4-sphere, which we will assume)
\beq{ad3}
N(S)=N_{3}-N_{4}+1.
\eeq
This is the number of independent Bianchi identities we can write
down.  By Euler's relation for the 4-sphere it leaves precisely
$N_{1}-N_{{0}}+1$ independent plaquettes, and the Jacobian for
changing from independent gauge fields to plaquette variables is one
\cite{halpern}:
\beq{ad4}
\int \prod_{l \in T} \d A_l \prod_{l'\in ST} \del(A_{l'}) \; 
F(P(A)) = 
\int \prod_{t} \d P(t) \prod_{s \in S} \del(\sum_{t\in s} P(t)) 
\;F(P).
\eeq
In eq.\ \rf{ad4} $ST$ denotes a spanning tree in $T$ and $S$ denotes
the set of $N(S)$ independent two-dimensional surfaces where the
Bianchi identities are enforced.  $l$ and $t$ are respectively links
and triangles of the manifold and we assume that the definition of the
positive orientation of links and triangles was uniquely chosen.
Notice that in this implementation each elementary gauge loop
(plaquette) has a length three, however each gauge potential
contributes to a number of neighbouring triangles which is a
non-trivial local geometric characteristic of the system.

Similar discussion can be made in the $T^*$ implementation.  The dual
plaquettes constructed from a $D$-dimensional triangulation can be
labelled by the $(D-2)$-dimensional sub-simplices in the
$D$-dimensional triangulation, which are ``encircled'' by the
associated dual plaquette. In four dimensions we shall label dual
plaquettes by the triangles they encircle.  The geometric properties
of the two gauge-invariant objects $P(t_i)$ and $P^*(t_i)$ are however
different: The length of the plaquette $P^*(t_i)$ is now $o(t_i)$,
where $o(t_i)$ is the order of the triangle $t_i$, or the number of
four-dimensional simplices sharing this triangle. As before we have to
choose a positive orientation, and the plaquette variable is unchanged
when the corresponding loop is rotated and changes sign, when it is
inverted. Each gauge potential $A^*_{ij}$ contributes to exactly four
dual plaquette variables.  Like in the $T$ implementation we can
parametrize the system by the plaquette variables $P^*(t_i)$,
satisfying a set of the Bianchi identities.

To this point we did not specify the model we want to consider.  Let
us start with the $T^*$ implementation.  The plaquette variables are
the sums of gauge fields associated with the oriented (dual) links
`encircling' the triangles $t_i$ in $T$:
\beq{2.0}
P^*(t_{i})= \sum_{\a} A^*(l^*_{\a}),
\eeq
where the orientation of the (dual) links follows that of the
plaquette and as discussed above $t_{i} \in T$ can be viewed as dual
to the plaquette. To keep the discussion general let us at this point
assume only that the gauge action is Gaussian and assumes a form:
\beq{2.1}
S^*_{g}(T,A^*) = \sum_{i} \b_i P^{*2}(t_{i}),
\eeq
where $\b_i$ are positive, local, geometry-dependent coefficients.  We
shall argue below what seems to be the most natural choice of $\b_i$.

With this notation the partition function can be written as
\beq{2.1a}
Z = \sum_{T} \frac{1}{C_T} \e^{-\k^*_4 N_4(T) +\k^*_2 N_2(T)} 
\int \prod_{l^*} \d A^*_{l^*} \prod_{l'^* \in ST^*} \del (A_{l'^*})\; 
\e^{-S^*_g (T,A^*)},
\eeq
where $ST^*$ is a spanning tree in $T^*$.

We shall now construct the dual version of this model.
We can write 
\beq{2.2}
\exp(-\b_i P^{*2}(t_i)) =
\sqrt{\frac{1}{\pi\b_i}} \int dp(t_i) \; 
\exp\Bigl(-\frac{p^{2}(t_i)}{\b_i}+2ip(t_i) P^*(t_{i})\Bigr).
\eeq 
Using this integral representation of the plaquette, as 
well as the standard integral of the $\del$-function:
\beq{ad6} 
\prod_{l' \in ST^*} \del (A^*_{l'}) =
\int \prod_{l' \in ST^*} \frac{\d \a_{l'} \;\e^{2 i \a_{l'} A^*_{l'}}}{\pi} 
\eeq
the integration   
over the $A^*$-fields in $T^*$ can be performed, 
giving rise to a set of delta functions: 
\beq{2.3} 
\prod_{l^*} \pi \del( \a_{l^*}+\sum_{t_{i} \ni l'^*} p(t_{i})), 
\eeq
where the product is over the dual links $l^*$, and $\a_{l^*} =0$ if
the dual link $l^*$ does not belong to the chosen spanning tree.
Notice that the sum in the argument of each $\del$ function has
exactly four terms.  Performing the $\a_{l^*}$ integration in \rf{ad6}
eliminates $N_4(T)-1$ of the $N_3(T)$ $\del$-functions in
\rf{2.3}. The remaining $N_3(T)-N_4(T)+1$ $\del$-functions can be
viewed as the Bianchi identities for the $N_2(T)$ variables $p(t_i)$.
Since there are $N_2(T)-(N_3(T)-(N_4(T)-1))$ independent variables
$p(t_i)$, we can implement the Bianchi identities by introducing
$U(1)$ gauge fields $A_l$ living on the links of the triangulation
$T$, rewriting $p(t_i) = P(t_i) = \sum_{l\in t} A_l$.  Independent
gauge transformations can be done at vertices, and after gauge fixing
we are left with $N_1(T)-(N_0(T)-1)$ independent $A_l$ fields, i.e.\
precisely the same as the number of independent $p(t_i)$ fields.

We end up this discussion by writing down the identity valid for each 
triangulation:
\bea \label{dualexpr}
\int \prod_{l^*} \d A^*_{l^*} \prod_{l'^* \in ST^*} \del (A_{l'^*})\; 
\e^{-S^*_g (T,A^*)} &=& \pi^{-\frac{1}{2}N_2 
+ \frac{3}{2}N_4 +1}\; \e^{-\oh \sum_{t_i} \ln \b_i}  \\ \nonumber
&\times&\int \prod_{l} \d A_{l} \prod_{l' \in ST} \del (A_{l'})\; 
\e^{-S_g (T,A)}, 
\eea
where
\beq{dualaction}
S_g(T,A)= \sum_{t_i} \frac{P^2(t_i)}{\b_i}
\eeq
and $P(t_i)$ is the oriented plaquette variable \rf{plaq} expressed in
terms of the fields $A_l$ living on the links of the triangulation $T$.
For the partition function \rf{2.1a} we get
\beq{2.1e}
Z = \sum_{T} \frac{\pi}{C_T} \e^{-\k_4 N_4(T) 
+\k_2 N_2(T) -\oh \sum \ln \b_i} 
\int \prod_{l} \d A_{l} \prod_{l' \in ST} \del (A_{l'})\; 
\e^{-S_g (T,A)},
\eeq
where $ST$ is a spanning tree in $T$ and 
\beq{2.1c}
\k_4 = \k^*_4 - {3\over 2}\ln \pi,~~~{\rm and}~~~\k_2 =\k^*_2 -\oh \ln \pi.
\eeq 
Thus, apart from a shift $\k_4\to \k^*_4-\frac{3}{2} \ln \pi$ and 
$\k_2\to \k^*_2- \oh \ln \pi $, 
the two models are dual up to the weight factor $-\oh \sum \ln \b_i$. 

The choice of the action parameters $\b_i$ should be made at this
point.  One can argue that it is natural to choose $\b_i=1/o(t_{i})$,
where $o(t_{i})$ is the order of the triangle. In the $T^*$
implementation $o(t_{i})$ is also the length of the boundary of the
plaquette dual to $t_i$, and, assuming the plaquette to be flat, also
proportional to it's area.  Since $P^*(t_i)$ signifies the flux of the
plaquette dual to $t_i$, the field strength is proportional to
$P^*(t_i)/o(t_i)$, while the volume element associated with $t_i$
likewise is proportional to $o(t_i)$. This argument assumes that the
plaquette is a flat two-dimensional object with a constant field
strength. This needs not be the case, so another possible choice could
be $\b_i=1/o^{\a}(t_{i})$ with some power $\a$ different from $1$,
which leaves space for some non-trivial geometry of plaquettes.

The model for a single non-compact Abelian gauge field can readily be
generalized to $n_g$ copies. For a given triangulation, $T$, these
models will be non-interacting, but summing over all triangulations
one introduces interactions between the copies, mediated by the
geometry, as well as interactions between geometry and the matter
fields.  The final action we use can thus be written as
\beq{2.6}
S_g^* = \sum_{i,k} \frac{P^{*2}(t_i,k)}{o^\a (t_i)},
\eeq
where the index $k$ numbers the copies of Abelian gauge fields and takes
values $1,\ldots,n_g$. If $\a=1$ the model
is dual to the original AGM model  except for a measure factor 
\beq{2.6a}
\oh n_g \sum_t \ln o(t) 
\eeq 
for each triangulation and a shift in the gravitational 
and cosmological constants by
\beq{2.6b} \k_4 \to \k_4+\frac{3n_g}{2} \ln \pi~~~{\rm and}~~~ 
\k_2 \to \k_2 +\frac{n_g}{2} \ln \pi. 
\eeq
We call this version of the model the dual Abelian gauge model (DAGM).

We can make two remarks at this point:
\begin{itemize}
\item The whole derivation made explicitly use of the fact that the
topology of the manifold is that of a four-sphere. For different
topologies the duality relations become slightly more complicated and
new degrees of freedom appear, related to the possible topologically
non-trivial boundary conditions for the gauge potentials.
\item Similar discussion can be made in the three and the
two-dimensional cases. In three dimensions in the $T^*$ implementation
the dual of a plaquette is a link. The duality relates the gauge
theory on dual plaquettes to a massless bosonic theory, where the
bosonic field live on vertices, again up to a ultra-local measure
term.  The $T$ implementation will have it's dual in the form of a
massless scalar field living in the centers of the simplices. In two
dimensions the duality transformation permits to integrate out the
gauge field completely, leaving only the measure term.
\end{itemize}

\section{The algorithm}\label{algorithm}

The algorithm describing numerical simulations of the four-dimensional
simplicial gravity was described in many papers. Here we shall not
repeat details concerning the five geometric ``moves'' which became a
standard. There are however some peculiarities of the present
simulation, which make the updating in the gauge sector rather
non-trivial.

In order to perform a Monte Carlo simulation of the model we found it
convenient to work directly with the gauge invariant plaquettes rather
than the gauge fields themselves.  Occasionally we shall need the
gauge fields and we shall then reconstruct them from the plaquette
variables. This is necessary only in order to control round-off
errors, which eventually will lead to a violation of the Bianchi
identities.  The geometric ``moves'' create and destroy simplices. In
effect also the new gauge potentials, located at the interfaces
between simplices are created and destroyed. Since the gauge action is
Gaussian we shall use the heat bath method to generate the new
variables. In order to find the detailed balance condition it is
always necessary to compare the ``move'' and it's inverse.

The addition of the new gauge potentials does not create any problems
and can be handled in a way closely resembling that of the Gaussian
scalar fields. The procedure has two steps: in the first step we
decide if the move will be performed. If the answer is yes we proceed
to the second step, where the new gauge potentials are generated from
a (multi-dimensional) normal distribution.  In the process the old
plaquette variables are modified and new plaquette variables
appear. The Gaussian form of the gauge potential guarantees that both
the modified and new plaquettes will remain restricted in value. The
acceptance probability for the first step can be expressed in terms of
the gauge invariant plaquette variables. In the second step the values
of new gauge potentials are generated, but only the gauge invariant
information (the plaquette variables) is stored.

Situation may be quite different when we try the inverse operation
deleting some gauge potentials. If we would decide to store the values
of the potentials and simply subtract these values from plaquettes,
the resulting change in the gauge action could become arbitrarily
large. The reason is that only the gauge invariant combinations of the
gauge potentials are physical (and in the Monte Carlo simulation
restricted by the gauge action), while the gauge potentials themselves
may become large unless some form of the gauge fixing is
imposed. Introducing the gauge fixing (a spanning tree) is however a
global problem and therefore maintaining the gauge condition in the
numerical simulation, where the geometry is dynamical would be very
costly and impractical. We have developed a different approach, which
is again a modification of the heat bath algorithm. In the first step
we decide if the inverse move will be performed. In order to find the
acceptance we integrate over all possible gauge choices of the gauge
potentials to be deleted.  The resulting formula is local and
expressible in terms of the gauge invariant quantities. If the inverse
move is accepted we proceed to the second step, where the potentials
to be deleted are generated from the appropriate normal
distribution. After the inverse move only the gauge invariant
information is stored.

In the program the situation is slightly more complicated. In order to
maintain the detailed balance at each move we have to consider the
acceptance for both the move and it's inverse. Also the geometric
moves used in the four dimensional updating of geometry in general are
rather complicated and in each move some gauge potentials are created,
while other are destroyed. We shall report more details about the
updating procedure in the appendix.

\section{Observables and measurements}\label{obs} 

Basic observables in simplicial quantum gravity have already been
discussed in a number of articles. Since we are here interested in
gauge fields it would be natural to measure gauge invariant
observables. An obvious gauge invariant quantity is the correlator $
\la F^2_{\m\n}(r) F^2_{\l\rho}(0)\ra$, where the average is over all
geometries with $S^4$ topology and with points separated a geodesic
distance $r$. The gauge field itself is just Gaussian, so the
propagator in flat space is trivially calculated.  In quantum
space-time, where we average over all geometries , it is not clear
precisely how the propagator will fall off with the geodesic
distance. In fact, the motivation for this study was to test if the
gauge fields had a stronger interaction with geometry than the scalar
fields investigated so far. In \cite{abj} it was shown how to extract
the {\it connected} correlator for matter fields coupled to simplicial
quantum gravity. However, in order to obtain a non-trivial result of
interest for continuum physics it is first necessary to identify a
point in the combined coupling constant space of gravity and gauge
theory where we can take an interesting continuum limit. In the
following we concentrate on this necessary first step. Thus we study
the $\k_2,\k_4$ phase diagram of the theory for different values of
the parameter $\a$ in eq.\ \rf{2.6}. For a given value of $\k_2$ there
exists a critical point $\k_4$ where an infinite volume limit can be
obtained.  At this critical point one can measure geometric quantities
like the entropy exponent $\g$, which describes the distribution of
baby universes, and one can measure the fractal dimension of
space-time.  As is standard in the measurements of the fractal
dimension in simplicial quantum gravity studies, one operates with
two, in principle independent, definitions of the fractal dimension.
A ``short distance'' fractal dimension, $d_h$, measures how the volume
$V(r)$ of spherical balls of geodesic radius $r$ grows with increasing
$r$ as long as $r$ is much smaller than the average radius $R$ of the
universe:
\beq{xz1}
V(r) \sim r^{d_h}~~~{\rm for}~~~r \ll R.
\eeq
A ``cosmological'' fractal dimension is defined by the way 
``macroscopic'' distances scale with the cosmological constant
or with the average space-time volume. For instance 
\beq{xz2}
V(r) \sim r^{d_h-b} F\Bigl(\frac{r}{V^{1/d_H}}\Bigr),
\eeq
where $F(x) \sim x^b$ for $x \to 0$. In the four-dimensional gravity
models studied so far one has always found that $d_h\equ
d_H$. Although there exist models \cite{c-2} of fractal structures
where $d_h \!\neq \! d_H$, one expects that $d_H=d_h$ if a sensible
thermodynamic limit exists \cite{www,ajw}.

Since the transition between the branched polymer phase and the
crumpled phase of four dimensional gravity is characterized by the
appearance of singular vertices of very high order, we also study the
distribution of the order of vertices, links, and triangles in order
to characterize the various phases we observe.

\subsection{$\a =1$}

\begin{figure}[t]
\centerline{\epsfxsize=4.0in \epsfysize=2.67in \epsfbox{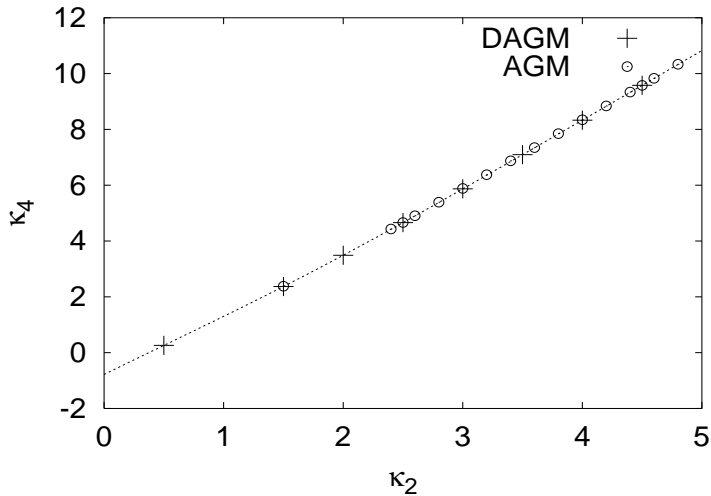}}
\caption{The $(\k_2,\k_4)$ diagram of the AGM and that of the dual model
(DAGM) studied in this paper. Error bars are smaller than the point
size and the line is only to guide the eye. $n_g=3$ and $N_4=4000$.}
\label{f:1}
\end{figure}
\begin{figure}[htb]
\centerline{\epsfxsize=4.0in \epsfysize=2.67in \epsfbox{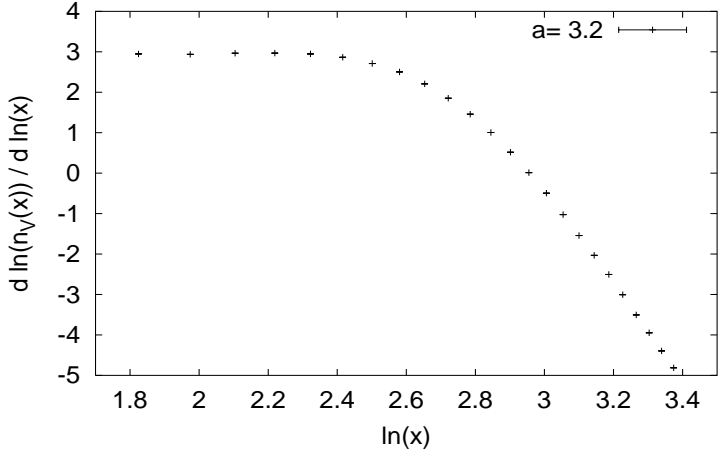}}
\caption{The logarithmic derivative of the two point function
$n_V(r)$. Here $x=(r+a)/V^{1/d_h}$ where $V\equiv N_4$. 
For a suitable choice of the shift
$a$ a power law defining $d_h$ can be obtained.}
\label{f:2}
\end{figure}

Since the model \rf{2.6} for $\a \equ 1$ is dual to the original AGM
model if we include the weight factor \rf{2.6a} and make the shift
\rf{2.6b}, we can as a calibration reproduce the $\k_2,\k_4$ diagram
of AGM. This is shown in fig.\ \ref{f:1} for the number of gauge
fields $n_g\equ 3$. We see perfect agreement.  As a general statement
we also see approximately the same values of $\g$ and $d_h$ as
observed in the AGM model, thus verifying the results obtained in
these investigations.  It is illustrated in Table\ \ref{t:1}.
\begin{table}[ht]
\begin{center}
\begin{tabular}{|l| l |l|}
\hline 
$\kappa_2$ & $\g$     & $d_h$    \\
\hline
 4.0  & -0.30(1) & 3.95 (5) \\
 4.5  & -0.12(1) & 3.57 (3) \\
\hline
\end{tabular}
\end{center}
\caption{Measurements of $\g$ and $d_h$  for
the DAGM model. $n_g=3$ and $N_4=4000$.} 
\label{t:1}
\end{table}
Our measurements were made on the $N_4=4000$ system with $n_g=3$. It
was therefore possible to measure only $d_h$ (and not $d_H$) by using
standard techniques \cite{ajw,aau,c-2}. The number $n_V(r)$ of
$4$--simplices at distance $r$ from a given $4$--simplex was measured
and a fit was made to the function $C(r+a)^{d_h-1}$ where $a$ is the
so--called ``shift''. $r$ is the ``$4$--simplex distance'' which is a
measure of the geodesic distance between $4$--simplices. The use of
the shift permits us to estimate $d_h$ even for this moderate size of
the system as can be seen in fig.\ \ref{f:2}. A slight modification of
the shift $a$ results in a much poorer fit for small $r$.

However, the important point is that for $\a = 1$ and {\it without}
the measure term one moves directly from the crumpled phase and into a
branched polymer phase. There is no trace of any crinkled phase and
the interaction between the gauge fields and geometry is as weak as
the interaction between scalar fields and geometry observed in
\cite{xxx}, provided that the gauge fields are coupled to simplicial
gravity in a way consistent with the way the scalar fields were
coupled to the geometry in \cite{xxx}, as discussed in detail in the
beginning of sec.\ \ref{themodel}.

In table \ref{tab1} we present the results of the computer simulations
for the number of gauge copies $n_g=3$ and $6$. The value of $\k_2$ is
chosen so large that one is well away from the crumpled phase. We
studied systems with volume $N_4= 4000, 8000$ and $16000$.  Our
results are summarized in Table\ \ref{tab1}.

\begin{figure}[htb]
\centerline{\epsfxsize=3.1in \epsfysize=3.1in \epsfbox{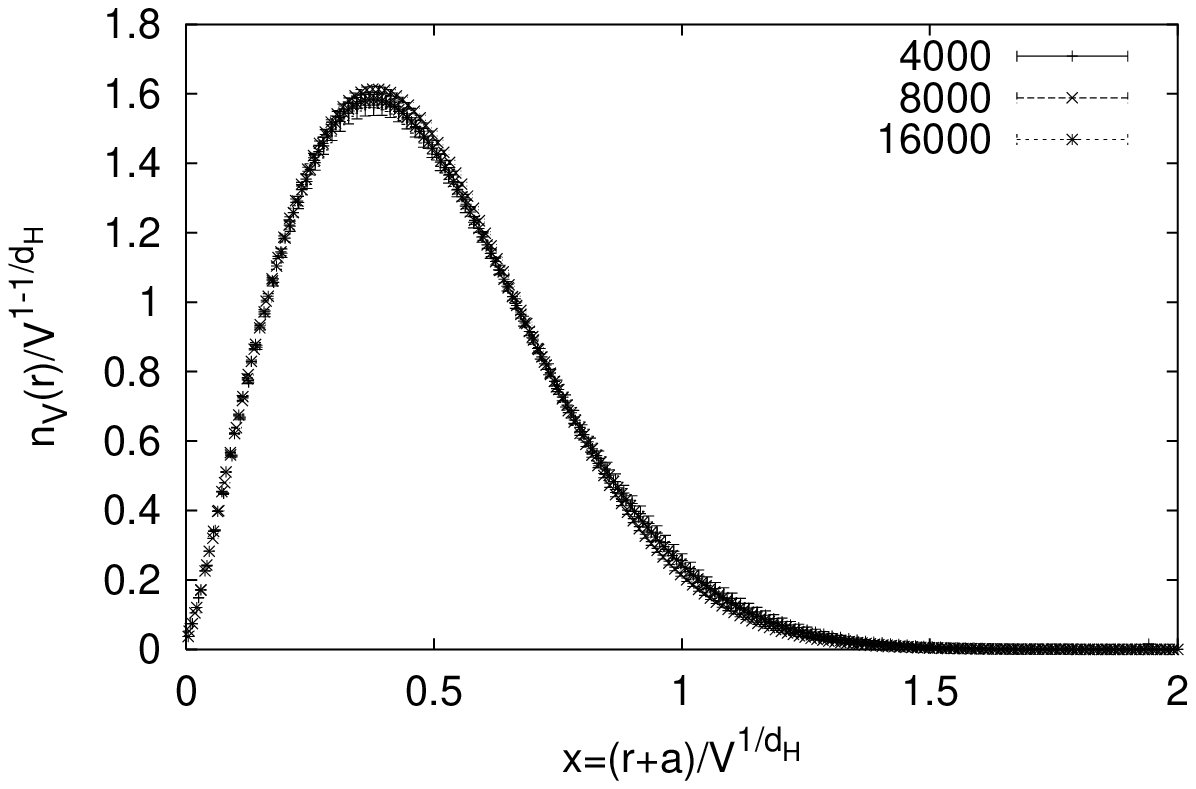}
            \epsfxsize=3.1in \epsfysize=3.1in \epsfbox{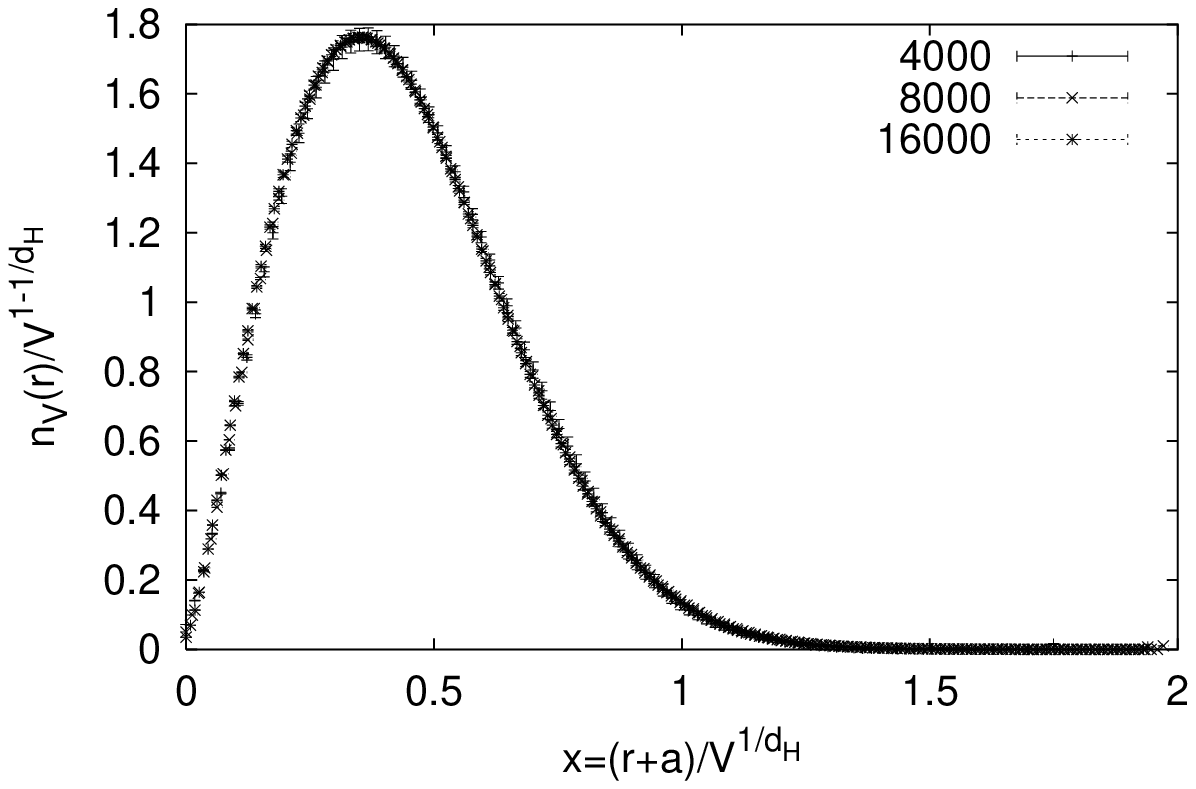}}
\caption{Calculation of $d_H$ using finite size scaling of
$n_V(r)$. The left figure corresponds to $n_g=3$, $\kappa_2=4.5$,
$d_H=2.03$  and the right one to  $n_g=6$, $\kappa_2=7.0$,
$d_H=2.05$.}
\label{f:3}
\end{figure}

\begin{figure}[htb]
\centerline{\epsfxsize=3.1in \epsfysize=3.1in \epsfbox{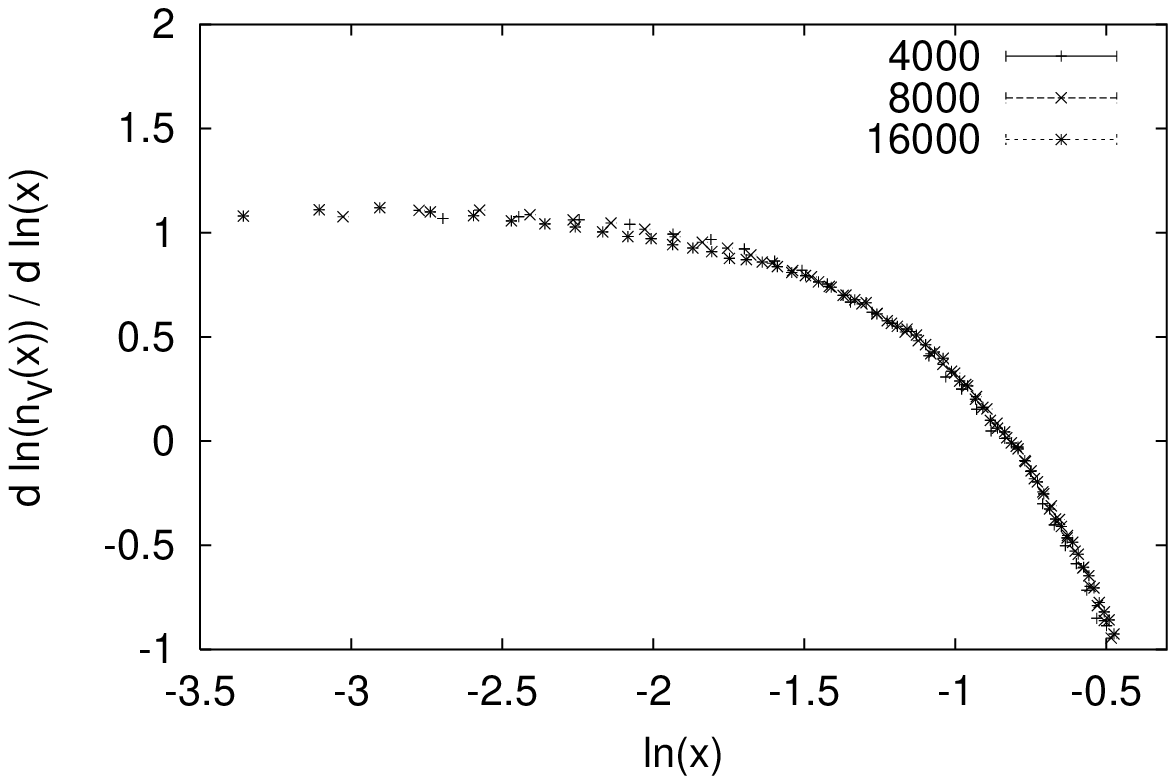}
            \epsfxsize=3.1in \epsfysize=3.1in \epsfbox{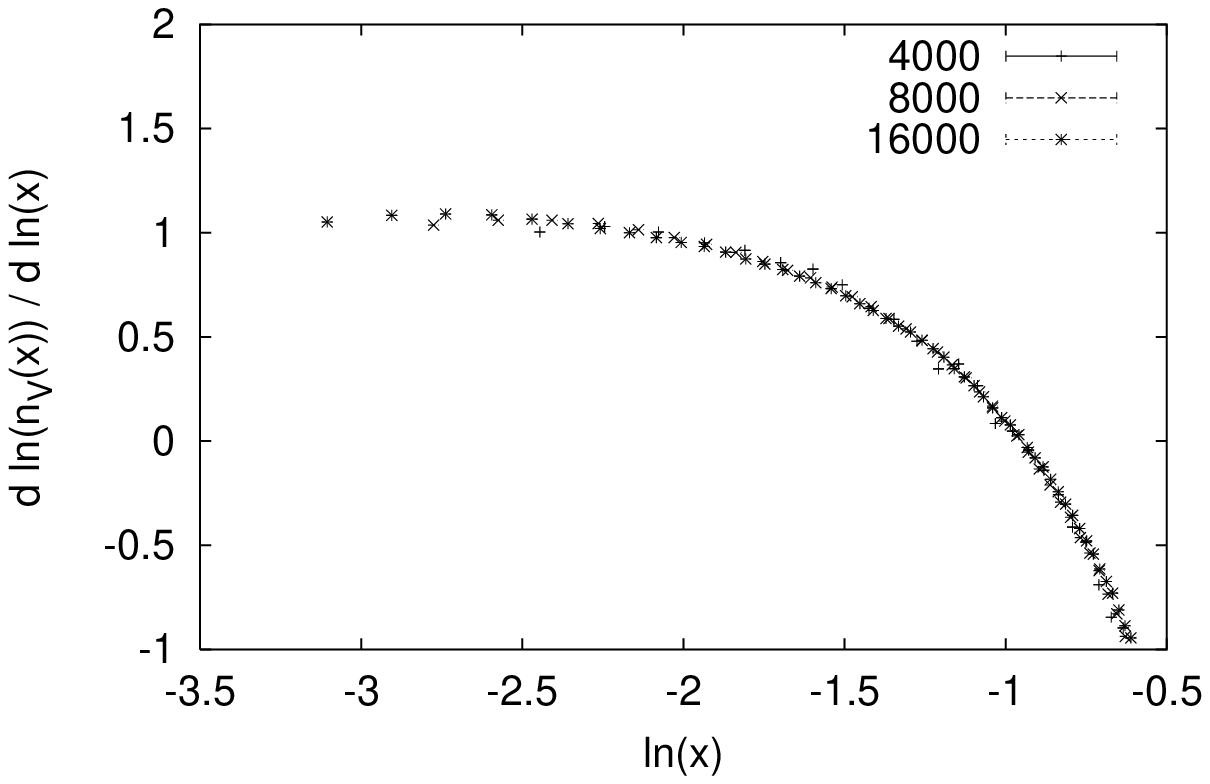}}
\caption{Calculation of $d_h$ for $n_g=3$, $\kappa_2=4.5$ (left) and
$n_g=6$, $\kappa_2=7.0$ (right). Here we use $x=(r+a)/V^{1/d_h}$. 
The shifts $a$ are $-0.5$ and $-1.5$ respectively. }
\label{f:4}
\end{figure}

\begin{table}[ht]
\begin{center}
\begin{tabular}{|l| l |l| l | l |l |}
\hline 
$n_g$ & $N_4$ & $\kappa_2$ & $\g$ & $d_h$  & $d_H$ \\
\hline
3 &  4000  & 4.5  & 0.49 (1) & 2.1 (2)&  2.03(7) \\
  &  8000  & 4.5  & 0.51(1)  &        &          \\
  &  16000 & 4.5  & 0.483(4) &        &          \\
\hline
6 &  4000 & 7.0   & 0.44 (1) & 2.2 (2)&  2.05(6) \\
  &  8000 & 7.0   & 0.47 (2) &        &          \\
  &  16000& 7.0   & 0.470(5) &        &          \\
\hline
\end{tabular}
\end{center}
\caption{Measurements of $\g$, $d_h$ and $d_H$ when $\a\equ 1$} 
\label{tab1}
\end{table}
In fig.\ \ref{f:3} and fig.\ \ref{f:4} we show the computation of
$d_H$ and $d_h$ from the scaling properties of $n_V(r)$. $d_H$ is
computed as described in \cite{aau,c-2} using the relation
$n_V(r)=V^{1-1/d_H}F_1((r+a)/V^{1/d_H})$. We determine $d_H$ and $a$
from the optimal ``collapsing'' of the $n_V(r)$ distributions. $d_h$
is obtained as for the DAGM model described above. We observe that the
expected scaling holds very well and that the values for $d_h$ and
$d_H$ are consistent with being the same.

\begin{figure}[htb]
\centerline{\epsfxsize=3.1in \epsfysize=3.1in \epsfbox{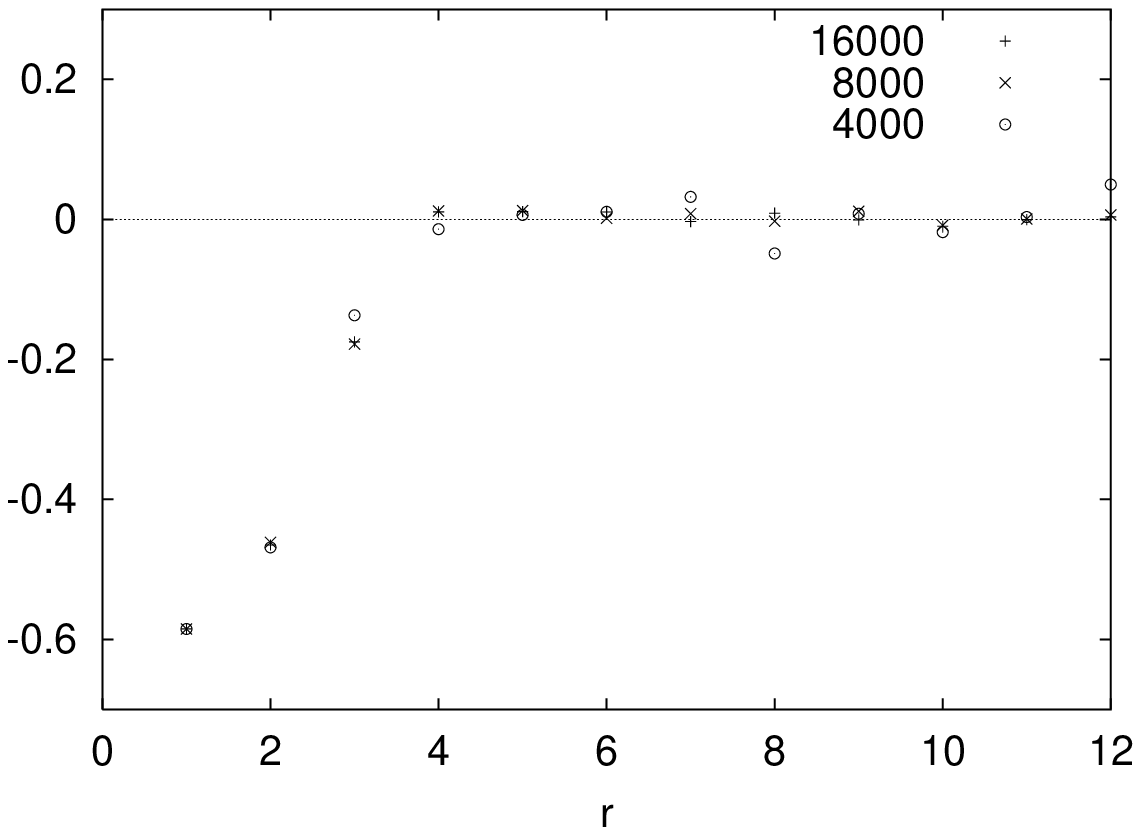}
            \epsfxsize=3.1in \epsfysize=3.1in \epsfbox{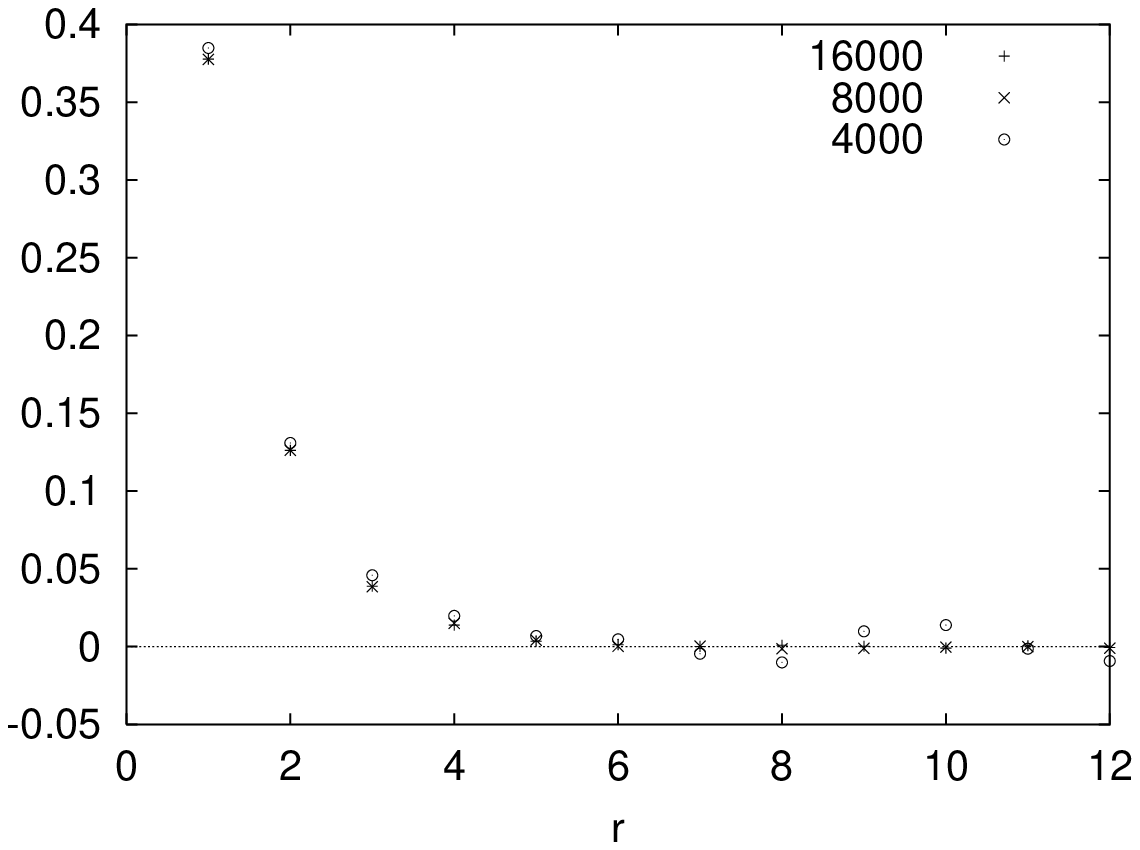}}
\caption{The connected curvature-curvature (left) and action-action
(right) correlation functions. $n_g= 6$ and $k_2= 7.0$. The $n_g= 3$
plots are falling on top of the $n_g= 6$.}

\label{f:5}
\end{figure}

We have also measured the connected part of curvature--curvature and
action--action correlators (we refer to \cite{abj}  for the details). 
One expects for two observables $A_i$, $i=1,2$ that
\beq{gA}
G^{1A_i}(r) = \vev{A_i} G^{11}(r+\delta_{A_i})\, ,
\eeq
and that the connected part of the correlator is given by
\bea\label{gAAc}
G^{A_1A_2}_c(r)&\equiv&   G^{A_1A_2}(r) - \vev{A_1} G^{1 A_2}(r +\delta_{A_1})
- \vev{A_2} G^{1 A_1}(r +\delta_{A_2})\\ \nonumber
&+&\vev{A_1}\vev{A_2} 
 G^{11}(r+\delta_{A_1}+\delta_{A_2})\, .
\eea
In the above formulas ``$1$'' refers to the unity operator and
$G^{11}(r)\equiv n_V(r)$.  In the case of curvature--curvature
correlator we determine the best $\vev{A_i}$ and $\delta_i$ by
collapsing $G^{1A_i}(r)$ with $G^{11}(r)$. We use these values in
\rf{gAAc} in order to produce the plot in fig.\ \ref{f:5}. Similarly
we proceed for the action--action correlator.  Observe that
curvature--curvature fluctuations are short range and independent of
the system size. The same is true for action--action correlators which
fall off at the same scale as the curvature--curvature ones. Identical
plots are obtained for the AGM model so we conclude that the measure
factor does not smoothen the configurations enough for the fields to
fluctuate at large scales.  The observed behaviour means that using
the scaled distances the correlations become ultra-local. Similar
pattern was observed in the crinkled and crumpled phases. This remains
in the qualitative agreement with the curvature-curvature correlations
measured in pure gravity \cite{bs}, where the definition of the
connected correlator was different.

The value for $\g$ as well as those of $d_h$ and $d_H$ are consistent
with those of ordinary branched polymers.  The branched polymer
interpretation is also corroborated by the fact that $\la N_0 \ra/N_4
\approx 1/4$ which is the upper kinematic bound expected from branched
polymers, plus the fact that we see no vertices of high order.

\subsection{$\a \neq 1$}

With the negative results of the above subsection in mind we turn to
the more general action \rf{2.6} with $\a \neq 1$ in a search for new
and potentially interesting fixed points from the point of view of
continuum physics. By performing the duality transformation for a
general $\a$ we obtain an action
\beq{3.1}
\sum_{t} \Bigl( \sum_{k=1}^{\n_g}o^\a(t) p^2(k,t)\Bigr) +\frac{\a n_g}{2}
\ln o(t).
\eeq
Thus there is a possibility that a choice of negative $\a$ might bring
us to a phase resembling or being identical to the crinkled phase,
since the negative weight factor will imitate the situation
encountered in the AGM model. We show the result of the measurements
for the number of gauge fields $n_g \equ 3$ and $\a\equ \mi 0.5$ and
$\mi 1$. The value of $\k_2$ is chosen such that we avoid the crumpled
phase.

The results of the measurements of $\g$ and $d_h$ is shown in Table\
\ref{t:2}.  There is a clear tendency for the values to $\g$ and $d_h$
to drift from the branched polymer values toward the crinkled values,
in accordance with expectations. Some limited statistics that we
obtained for the $\a=-1.5$ model suggest that $\g$ becomes even more
negative. We have taken this as evidence that we see not really new
physics with the modified weights, but rather an effective change in
measure, more or less in the same way as the original AGM model
differs from our model by measure term.
\begin{table}[ht]
\begin{center}
\begin{tabular}{|l| l |l|}
\hline 
$\a$ & $\g$     & $d_h$    \\
\hline
 -0.5  &  0.30(2) & 2.65 (3) \\
 -1.0  & -0.51(1) & 3.5  (2) \\
\hline
\end{tabular}
\end{center}
\caption{Measurements of $\g$ and $d_h$  for
$\a < 0 $. $n_g=3$, $N_4=4000$ and $\kappa_2=4.5$.} 
\label{t:2}
\end{table}

\section{Discussion}

Gauge fields contribute far more to the conformal anomaly than scalar
fields. Thus they are expected to play an important role in effective
models of four-dimensional quantum gravity where the (infrared)
dynamics is dictated by the conformal anomaly \cite{amz}. The
conjectured scaling behavior of these model are quite similar to the
observed (pseudo) scaling behavior observed in four-dimensional
simplicial quantum gravity, and this led to the expectation that gauge
fields might couple stronger to geometry than the scalar fields used
so far. So far, computer simulations have revealed only a weak
coupling between the scalar fields and the geometry, and the presence
of scalar fields have not led to any quantitative change in the phase
diagram of simplicial quantum gravity.

The first Monte Carlo simulations of simplicial quantum gravity
coupled to gauge fields seemed to be in accordance with the above
philosophy. Indeed, for the first time one observed a genuine
back-reaction of the matter fields on the quantum geometry.  However,
as already remarked in the original article \cite{bbkptt} and verified
in detail later \cite{bbkptt1}, major part of the interaction between
geometry and the matter could be accounted for by the term
\beq{4.1}
S_{eff}= \frac{n_{g}}{2} \sum_{t} \ln(o(t)),
\eeq
where $o(t)$ denotes the order of the triangle $t$. This term is an
ultra-local measure term and it is (as already remarked in
\cite{bbkptt}) unlikely that it contributes to the infrared dynamics
related to the conformal anomaly in the model proposed in
\cite{amz,amz1}. However, it still left us with the puzzle why gauge
fields seemingly interact so much stronger with geometry than ordinary
scalar fields.

In the work presented here we have verified the original results
obtained in the AGM model by an independent Monte Carlo simulation
working with gauge fields on the lattice dual to that of the of the
AGM model. In addition we have shown that there is {\it no} major
difference between the interaction of scalar fields with geometry and
the interaction of gauge fields with geometry. In fact, if the gauge
fields are coupled to geometry in a way consistent with the way the
scalar fields were coupled to geometry, the gauge and the scalar
fields have similar weak coupling to the geometry.  We have explained
how the weight factor \rf{4.1} arose in the transformation from gauge
fields living on the dual links of a triangulation to gauge fields
living on the links themselves.

Turning the arguments around, our results lead to the prediction that
scalar fields living on the vertices of four-dimensional
triangulations will couple as strongly to the geometry as the gauge
fields in the AGM model.

We extended that gauge field model coupled to geometry by modifying
the weight given to each plaquette, in order to explore further the
phase-space structure of matter coupled to geometry.  However, the
only new phase we could identify is presumably identical to the
crinkled phase already found in the AGM model. The weight generated by
the transformation from the dual links to the ordinary links of a
triangulation gave support to this interpretation.

\section*{Appendix}

All moves require modification of some plaquettes by introducing new
dual links into plaquettes or by destroying the dual links. Finding
probabilities for the moves resembles the problem of the Gaussian
scalar fields, however the gauge invariance of the action requires
some important changes to maintain the detailed balance and at the
same time to keep the acceptance reasonable.  In this section we shall
discuss the case of one gauge potential and the parameter $\alpha =
1$.\

Let us first describe the process of introducing new degrees of
freedom.  For simplicity let us discuss only the situation of the
move, where a new vertex is added in the center of a simplex. There
are now 10 new gauge potentials. We consider only the part of the
gauge action involved in the move.  Before the move is attempted we
have
\begin{eqnarray}\label{up0}
\exp (-S_A(p_i)) =
\exp (-\sum_i^{10} \b_i p_i^2),
\end{eqnarray}
where $\b_i=1/o(t_i)$ and $p_i$ are the {\em external plaquette}
variables.  The modified action becomes
\begin{eqnarray}\label{up}
\exp (-S_B(p_i,x_i)) =
\exp (-\sum_i^{10} \b'_i (p_i+x_i)^2 -\b' \sum_{i,j} x_i M_{ij} x_j)
\end{eqnarray}
where $x_i$ are new fields, $\b'_i$ are the new weights, typically
$\b'_i=1/(o(t_i)+1)$, the matrix $M_{ij}$ contains the geometric
information about ten new {\em internal plaquettes} formed in the
move.  These new plaquettes $\Pi_k$ can be written as
\begin{eqnarray}
\Pi_k = \sum U_{kj}x_j\\
M_{ij}=\sum_k U_{ki}U_{kj}
\end{eqnarray}
where $U_{ki}$ are $\pm 1$ or zero reflecting the orientation of the
$x_i$ fields. The internal plaquettes have always the order three, so
$\b'=1/3$ typically.  When performing the move we shall have to
\begin{itemize}
\item decide whether we make the move or not and
\item if yes -- generate the new fields with the distribution resulting from 
(\ref{up}).
\end{itemize}
Let us define
\begin{eqnarray}
{\cal N}_B(p_i)&=&\int \prod dx_i \exp(-S_B(p_i,x_i)) =\nonumber \\
&=& \sqrt{\frac{\pi^{10}}{{\rm det}({\bf Q})}}
\exp(-\bar{S}_B(p_i)),\label{up1} \\
Q_{ij}&=& \b'~ M_{ij} + \b'_i~ \del_{ij}, \nonumber
\end{eqnarray}
where $\bar{S}_B(p_i)$ is the {\em global minimum} of $S_B(p_i,x_i)$
with respect to $x_i$. This minimum can of course be explicitly
calculated in terms of $p_i$ and the matrix ${\bf Q}^{-1}$:
\begin{eqnarray}
\bar{S}_B(p_i)=\sum_i \b'_i p_i^2 -\sum_{ij}p_iQ^{-1}_{ij}p_j.
\end{eqnarray}

The quantity $w_B(p_i,x_i)=\exp(-S_B(p_i,x_i))/{\cal N}_B(p_i)$ is a
normalized Gaussian distribution of the $x_i$ fields. Let ${\cal P}_B=
\pi_B{\cal N}_B$ be the probability that the move will be performed
and, if accepted, the new fields be generated with this distribution.
It is easy to check that the transition probability $P(A \to B)$
satisfies
\begin{eqnarray}\label{up2}
\exp(-S_A)P(A \to B)&=& \exp(-S_A){\cal P}_B w_B(p_i,x_i)\\ \nonumber
 &=& \pi_B \exp(-S_A - S_B).
\end{eqnarray}  
The form (\ref{up2}) will be useful in order to discuss the detailed
balance condition.  The important point is that the only non-trivial
quantity, ${\cal N}_B(p_i)$ can be calculated also in the new
configuration \{$P_i = p_i+x_i,\Pi_k$\}. Using (\ref{up1}) we can
express $\bar{S}_B$ as a {\em global minimum} of $\sum_i \b'_i
(P_i+y_i)^2+\b'\sum_k(\Pi_k+\sum_j U_{kj}y_j)^2$ with respect to the
shifted variables $y_i$. The only difference is the shift of variables
and the value of the minimum is unchanged. This property is very
important, because we shall have to calculate ${\cal N}_B$ both for
the move and it's inverse.

Let us now discuss the inverse move: we destroy 10 internal plaquettes
$\Pi_k$ and modify values of 10 external plaquettes, eliminating 10
gauge fields produced before. The naive proposition would be to store
the gauge fields $x_i$ and simply to delete them. The gauge fields
however have no gauge invariant values so the transition performed
this way would depend on the particular gauge choice - in fact even if
plaquette variables are small (because of the Gaussian weights) the
gauge fields may become large and in effect the transition may be
blocked completely.

In the program we use a different approach. Rather than storing the
gauge fields we try to reconstruct them when needed. When trying to
delete fields $x_i$ we have first to decide what gauge choice we make.
In other words we shall study the change $P_i \to p_i=P_i-x_i$ where
$x_i$ satisfy the set of gauge--invariant constraints $\sum
U_{ki}x_i=\Pi_k$.  Notice that not all these constraints are
independent because of the Bianchi identities between the ten
plaquettes $\Pi_k$.  Only six of them have to be used, reducing the
number of independent degrees of freedom to four.  As before let us
consider only the part of the action engaged in the move.  Before the
move is attempted we have
\begin{eqnarray}\label{down0}
\exp(-S_B(P_i,\Pi_k))=\exp(-\sum_i^{10}\b'_i~P^2_i-\b'~\sum_k^{10}\Pi^2_k)
\end{eqnarray}
 After the
move we have: 
\begin{eqnarray}\label{down}
\exp(-S_A(P_i,x_i))=\exp(-\sum_i^{10} \b_i (P_i-x_i)^2),
\end{eqnarray}
As before we shall have to
\begin{itemize}
\item decide if we perform the move and
\item if yes, perform it, {\em i.e.} choose some values of $x_i$ following
from (\ref{down}).
\end{itemize}
Let me define as before:
\begin{eqnarray}
{\cal N}_A(P_i,\Pi_k)&=&\int \prod_i^{10} dx_i
\prod_k^6 \delta(\Pi_k - \sum U_{kj}x_j)
\exp(-S_A(P_i,x_i), \nonumber \\
 &=&\sqrt{\frac{\pi^4}{{\rm det} {\bf N}}}\exp(-\bar{S}_A(P_i,\Pi_k)),
\label{down1}
\end{eqnarray}
where the matrix ${\bf N}$ depends only on $\b_i$ and
$\bar{S}_A(P_i,\Pi_k)$ is the {\em conditional minimum} of
$S_A(P_i,x_i)$ with gauge--invariant set of conditions imposed on the
$x_i$ fields. The integration in (\ref{down1}) can be viewed as
integration over all possible gauge choices for the fields $x_i$. It
is simple to give the explicit formula for $\bar{S}_A$ in terms of
$P_i,\Pi_k$ and the matrix ${\bf N}^{-1}$. We shall not write it
here. Let us however note that ${\cal N}_A(P_i,\Pi_k)$ can be also
calculated in the new configuration $\{P_i,\Pi_k\} \to \{p_i,0\}$
using a simple relation ${\cal N}_A(P_i,\Pi_k) = {\cal N}_A(p_i,0)$.

The quantity $w_A(P_i,\Pi_k,x_i)=\exp(-S_A(P_i,x_i)/{\cal
N}_A(P_i,\Pi_k)$ is a normalized probability for $x_i$ satisfying the
gauge-invariant constraints.  As before let us choose the probability
${\cal P}_A = \pi_A {\cal N}_A(P_i,\Pi_k)$ to be the probability of
performing the move. If accepted we generate four independent $x_i$
from the distribution $w_A$. The remaining six can be calculated using
the constraints. We can check that the transition probability $P(B \to
A)$ satisfies
\begin{eqnarray}\label{down2}
\exp(-S_B)P(B \to A) = \pi_A \exp(-S_A - S_B).
\end{eqnarray}

Equations (\ref{up2}) and (\ref{down2}) can be used to write the
detailed balance condition:
\begin{eqnarray}
\exp(-S_A)P(A \to B) = \exp(-S_B)P(B \to A) \nonumber
\end{eqnarray}
meaning
\begin{eqnarray}
\frac{\pi_B}{\pi_A}&=&1 \nonumber \\
{\rm or}~~\frac{{\cal P}_B}{{\cal P}_A}
&=&\frac{{\cal N}_B}{{\cal N}_A}. \label{dbc}
\end{eqnarray}
As usual, this form of the detailed balance condition suggests that in
order to maximize the acceptance we need to know both ${\cal N}_B$ and
${\cal N}_A$.The generalization to the case of more gauge fields and
$\alpha \ne 1$ is trivial.

The same method can be used to calculate the weights for other
moves. The general rule is that we need to consider at the same time
both the move and it's inverse.

\section*{Acknowledgements}
J.J. would like to acknowledge a partial support by the Polish Government
Project (KBN) grants 2~P03B~00814 and  2~P03B~01917. J.A., K.A. and J.J 
 thank MaPhySto, Centre for Mathematical Physics and Stochastics, funded 
by a grant from The Danish National Research Foundation, for financial support.

\end{document}